\documentstyle[epsf,aps,twocolumn]{revtex}

\begin{document}

\author{G. Camelo-Neto e S. Coutinho \\
 Departamento de F\'\i sica\\
Universidade Federal de Pernambuco\\
CEP 50670-901, Recife, Brazil}
\title{Dynamical Model for Virus Spread}

\maketitle

\centerline{\bf Abstract}

\begin{abstract}

 The steady state properties of the mean density population of infected cells
in a viral spread is simulated by a general forest fire like cellular automaton
model with two distinct populations of cells ( permissive and resistant ones)
and studied in the framework of the mean field approximation. Stochastic
dynamical ingredients are introduced in this
model to mimic cells regeneration (with probability {\it p})
and to consider infection processes by other means than contiguity (with
probability {\it f}). Simulations are carried on a $L \times L$ square
lattice considering the eigth first neighbors. The mean density population of
infected cells ($D_i$) is measured as function of the regeneration probability
{\it p}, and analized for small values of the ratio {\it f/p } and for distinct
degrees of the cell resistance. The results obtained by a mean field like
approach recovers the simulations results. The role of the resistant parameter
$R$ ($R \geq 2)$ on the steady state properties is investigated and discussed
in comparision with the $R=1$ monocell case which corresponds to the {\em self
organized critical} forest fire model. The
fractal dimension of the dead cells ulcers contours were also estimated and
analised as function of the model parameters.\footnote{Presented at the
International Conference on Future of Fractals, Aichi, Japan, 25-27 july 1995.
To be published in Fractals 95.}

\end{abstract}

\section{Introduction}

In this work we study the steady state properties of a simple discrete
dynamical stochastic cellular automaton model describing the virus spread.
An earlier deterministic version of the present model was proposed by
Landini et al~\cite{Landini} in attempt to study the morphology of Herpes
Simplex Virus
(HSV) corneal ulcers. In this latter model two distinct populations of
epithelial cells were considered accordingly with the degree of permissivity
to infection by HSV. The susceptibility to viral infection by contiguity
were distinguished by assuming that the permissive cells become infected by
the presence of one or more infected neighbors while for the resistant ones
at least $R$ $(R\geq 2)$ infected neighbors cells are required to spread
de virus. Landini et al~\cite{Landini} considering a square lattice monolayer
cellular
automaton model (taking into account the first eight nearest neighbors)
showed that the HSV ulcers can be quantitatively well characterized by the
fractal dimension of their contour. Furthermore, they shed some light to
understand that the complex shape and evolution of the HSV ulcers may depend
on intrinsic characteristics (susceptibility to infection) and topological
distribution of the ephithelial cells. For instance, when the resistance
parameter $R=5$, Landini's model exhibit a dramatic changes from
dendritic to amoeboid morphology when the concentration of
permissive cells reaches the percolation threshold.

The aim of this work is to investigate the dynamical features of the {\em HSV}
ulcers evolution, that were well described by the {\em static} model proposed
by Landini {\em et al}~\cite{Landini}. We are mainly concern with the overall
distribution of ulcers in the steady state that should occurs after a primary
infection. Nowadays it is well known that many viruses have evolved mechanisms
to avoid the immune system, like e.g. adenovirus, murine and {\em HSV}. For
instance Hill {\em et al}~\cite{Hill} describe a new mechanism by which {\em
HSV} may evade the immune control. For those viruses after a primary infection
the virus enters a latent phase and can reactivate later on at any time due to
other factors. Therefore we introduce stochastic dynamical ingredients to the
model proposed by Landini {\em et al}~\cite{Landini} to mimic regeneration of
the ephithelial tissue and to consider re-infection processes due to latent
virus phase. With these ingredients
our model can also describe a forest fire phenomena with two distinct
species of trees with different degrees of resistance to burning. Actually,
the particular case with $R=1$ (one specie of tree) recovers the
Self-Organized Critical forest fire model with lightning probability
introduced by Drossel and Schwabl~\cite{Drossel}.

In the present work, two methods are employed to study the role of the
resistance parameter $R$ on the steady state properties of the dead cell
population (destroyed trees or empty sites in the counterpart forest fire
model):\ (a) an analytical mean field approach and (b) a numerical
simulation. In section II we describe our cellular automaton model and
develops the mean field approach. Section III is devoted to discuss the
numeric simulation procedure. The results and discussions are given on section
IV.

\section{The Model and the Mean Field Approach}

Our stochastic discrete cellular automaton is defined on a square lattice
with $L^2$ sites, the sites being occupied by a permissive cell, a resistant
cell, an infected cell or a dead cell. The system is parallel updated and
governed by the following rules during one time step:

\begin{description}
\item (a) a permissive cell becomes infected by contiguity if exist one or more
infected neighbor cells on it environment.

\item (b) a resistant cell becomes infected by contiguity if exist R or more
infected neighbors cells on it environment.

\item (c) an infected cell becomes a dead cell.

\item (d) a living cell (permissive or resistant) may becomes infected with
probability {\it f} if there are no enough infected neighbors on it
environment.

\item (e) a dead cell may regenerates with probability {\it p}, a fraction {\it
q} being permissive cells and (1-{\it q}) being resistant cells.

\item (f) the cell environment is defined by its eight nearest neighbors.
Therefore R ranges from 2 to 8.

\end{description}

 To study the mean field behavior of the system we will assume that
the system size {\it L} is large enough to prevent finite-size effects, and
that the system reaches a steady state after a transient period from
arbitrary initial conditions.\ We also assume that this steady state depends
only on the model parameters and is characterized by mean values of the
densities of the permissive cells $D_p,$ of the resistant cells $D_r,$ of
the infected cells $D_i$ and that of the dead cells $D_{d}$. These densities
are constrained by normalization condition:

\begin{equation}
D_p+D_r+D_i+D_d=1
\label{normalization}
\end{equation}

Now let's consider the variation of these densities within one time step.\
The rate equation for dead cells is clearly given by

\begin{equation}
\Delta D_d=D_i-pD_d,
\label{deltaDd}
\end{equation}

\noindent while for the permissive cells we have

\begin{equation}
\Delta D_p=qpD_d-[\Theta +f(1-\Theta )]D_p
\label{deltaDp}
\end{equation}

In Eq.~\ref{deltaDp}, the first term gives the mean number of new regenerate
permissive
cells and the second reads for the permissive cells becoming infected on the
next time step (rules (a) and (d)), $\Theta =1-(1-D_i)^8$ being the
probability for a given permissive cell to have one or more infected
neighbors. By analogy for the rate of resistant cells we have

\begin{equation}
\Delta D_r=(1-q)pD_d-[\Phi _R+f(1-\Phi _R)]D_r
\label{deltaDr}
\end{equation}

\noindent
where $\Phi _R$ is the probability that a given resistant cell is
surrounded by $R$ or more infected cells, that is

\begin{equation}
\Phi _R=\sum\limits_{n=R}^8\left(^{8}_{n}\right)D_i^n\left( 1-D_i\right) ^{8-n}
\label{Phi}
\end{equation}

\noindent
Finally the rate of the infected cells is given by,

\[ \Delta D_i=\left( \Theta D_p+\Phi _RD_r\right) + \]
\begin{equation}
\left[ \left( 1-\Theta \right)D_p + \left( 1-\Phi _R\right) D_r\right] f - D_i
\label{deltaDi}
\end{equation}

\noindent
 In the above equation the first term describes the number of cells
infected by contiguity in one time step, the second gives the ones infected
by other means and the third clearly stands for the number of infected cells
that will become dead in the next time step.

The steady state is characterized by constant mean densities of all
populations of cells, that is,

\begin{equation}
\Delta D_r=\Delta D_p=\Delta D_i=\Delta D_d=0
\label{equilibrio}
\end{equation}

\noindent In that conditions we obtain from Eqs. (\ref{deltaDd}-\ref{deltaDr})
that,

\begin{equation}
D_d=\frac{1}{p}D_i
\label{Dd}
\end{equation}

\begin{equation}
D_p=\frac{qD_i}{\Theta +f\left( 1-\Theta \right) }
\label{Dp}
\end{equation}

\begin{equation}
 D_r=\frac{\left( 1-q\right) D_i}{\Phi _R+f\left( 1-\Phi _R\right) }
\label{Dr}
\end{equation}

Now substituting these expressions onto the normalization conditions
(Eq.(\ref{normalization})) we finally obtain an equation for the infected cells
populations as
function of the model parameters (provide $f,p\neq 0$) given by,

\begin{equation}
D_i\left[ 1+\frac 1p+\frac q{\left( 1-f\right) \Theta +f}+\frac{1-q}{
\left( 1-f\right) \Phi _R+f}\right] =1
\label{Di}
\end{equation}

This equation, which express the mean field behavior of the system, can be
solved numerically, and the figures for $D_i$ can be used to obtain the
others steady state populations given by Eqs. (\ref{Dd}-\ref{Dr}).

\section{Numerical Simulation Procedure.}

We carry on numerical simulations of the present model by considering an $%
L\times L$ square lattice with periodic boundary conditions. The initial
state is prepared by randomly distributing permissive cells with probability
$0.4$ (arbitrary) and resistant cells with probability $0.6$. An infection
cell is triggered on central site of the lattice. After a transient time
interval (time steps) of the order of $L$ we measured all kind of cells
densities and then average these values over the next $L$ consecutive
configurations. During all processes infected cells were triggered randomly at
$1/f$ time steps and regenerated cells were randomly allowed at $1/p$ time
steps. As far as we are interested to investigate the possibility of
occurrence of {\em self-organized critical }steady states we fixed the model
parameters such that $1/f\gg 1/p\sim L$ . This is a necessary condition to
allow that all living cells of an infected cluster become infected before
new regenerated cells appears on the clusters ends. Note that $L/2$ is the
maximum time interval for infection to spread in a large living cells
cluster. The {\em fractal dimension} of the contours of dead cells clusters
were also
estimated by the box counting method and averaged over all simulations.

\section{Results and Discussion}

Since we are mainly interested in the properties of the dynamic steady state
of the virus spread we direct our attention to the mean density of the
population of infected cells $D_i$ or mean density of fire in the
counterpart forest fire model. In figure 1, we show the dependence of $D_i$
as function of the regeneration probability $p$ obtained by simulation of
the model on a $L=100$ square lattice with the degree of resistance varying
from $R=1$ to $8$, for the ratio $f/p=0.1$ and $q=0.6$. We notice that the
curve for $R=2$
looks qualitatively similar to the one obtained for $R=1$ ({\em SOC forest
fire model}) and is very distinct to the ones obtained for $R\geq 4$. Those
latter curves show an universal like dependence for lower values of $p$.
Furthermore, the plot for $R=3$ exhibit a crossover behavior between the $%
R\leq 2$ and $R\geq 4$ curves as $p$ varies from lower to upper values. Now we
make comparison with the mean field like results obtained from Eq.~\ref{Di}. In
figure 2 we display the results obtained by the mean field procedure recovering
the ones obtained by simulations. For $R=1$ and $2$ the mean field curves are
in qualitative agreement of the one obtained by simulations, while for $R=3$
one also observes the cross-over between $R \leq 2$ and $R \geq 4$ behaviors
with a very sharp changes at small but definite value of $p$. For $R \geq 4$
the agreement between both approaches is quantitatively established.

In figures 3 and 4 we present the behavior of the density of dead cells
(ulcers) as function of the regeneration probability as obtained by simulation
and mean field approaches, respectively. The same qualitative and quantitative
behavior observed for the density of infected cells (figures 1 and 2) occurs
for the dead cell density. To explore the crossover behavior observed for the
$R=3$ case, we plotted in figure 5 the density of infected cells $D_{i}$
against the regeneration probability $p$ for several values of the parameter
$(f/p)$ obtained by the mean field approach (Eq.~\ref{Di}). We observe that the
mean field solution (that is independent of
finite size effects) changes dramatically as $p$ approaches from a certain
value which is $f/p$ dependent. As $p$ decreases and approaches to zero the
condition $%
f/p\ll 1$ is no longer fulfilled and the SOC state disappears~\cite{Drossel}.
This means
that clusters of resistant cells remains alive for long time intervals
stopping the virus spread. In figure 6, we compare $R=3$ case obtained by both
methods by showing $D_{i} \times p$ plot for small values of $p$. This plot
indicates that for $R=3$ and for very small values of $p$ and for a fixed ratio
$f/p$ the mean field predicted behavior is confirmed by simulations, even for
finite small lattices. Therefore the expected {\em SOC} state for the dynamical
behavior of the clusters of dead cells (ulcers), similar to the one observed in
the forest fire model $(R=1)$~\cite{Drossel} for small values of $p$ should
occurs only for $R=2$ case.

We have also investigated the dependence of the mean density of infected
cells $D_i$ against the variation of the parameter $q$ (the fraction of
permissive regenerate cells) for a fixed value of $p$. This is shown in
figure 7 from a simulation on a $L=200$ square lattice and for $f/p=0.1$ and
$p=0.6$ . One observe the crossover behavior for the $R=3$ case as $q$
increases from lower to upper values.

Finally we have investigated the dependence of the {\em fractal dimension}
$D_{F}$ of the countour of the dead cells clusters as a function of the model
parameters. The clusters were generated by simulations and after the transient
time interval ( of order of $L$ ) the {\em fractal dimension} of ulcers
contours were calculated by using the {\em box counting} method and averaged
over the next $L$ equilibrium configuration.

In figure 8 we show $D_{F}$ as function of small values of ($f/p$) obtained for
$p=0.1$ and for $R=2,3$ and $4$. The figures for $R>4$ as quite similar to the
ones for $R=4$. We notice that for the models with resistant cells with
$R\geq3$ the fractal dimension of the ulcers are very sensitive to small values
of the ($f/p$) parameter. Under these conditions the densities of dead cells is
of order of $0.2$ (see figure 3) indicating that a steady state regime is
achieved where small ulcers of very rough contour remain dynamically stable.
For ($f/p$) close to $0.01$ the ulcers frontiers are less rough  for $ R >2 $
with $D_{F} \sim 1.4$. This figure is close to the one estimated by box
counting digitalized ulcers images of dentritic small real ulcers (Feret's
diameter ranging from 1.6 t0 3.2 mm)~\cite{Landini} . On the other hand if the
cell resistance is low ($R\leq2$) this latter picture does not hold and the
ulcers is allowed to spread over the whole tissue with high density and with
$D_{F}\sim2$.

In conclusion, we show that our present dynamical model indicates that the
degree of resistance of the cell to infection by contiguity should plays an
important role on the ulcer evolution.\\

\noindent
{\bf Ackowledgements:}

We acknowledge the financial support received from CNPq, FINEP and CAPES
(Brazilian granting agencies). One of us (G.\ C.\ N.) is gratefull to CNPq for
the Schorlarship for Scientific Initiation and the other (S.\ C.) acknowledges
to FACEPE (Pernambuco State granting agency) for financial support to present
this paper in International Conference on {\em Future of Fractals}, Aichi,
Japan, 1995.


\begin{figure}
\begin{center}
\leavevmode
\vbox{%
\epsfxsize=6cm
\epsffile{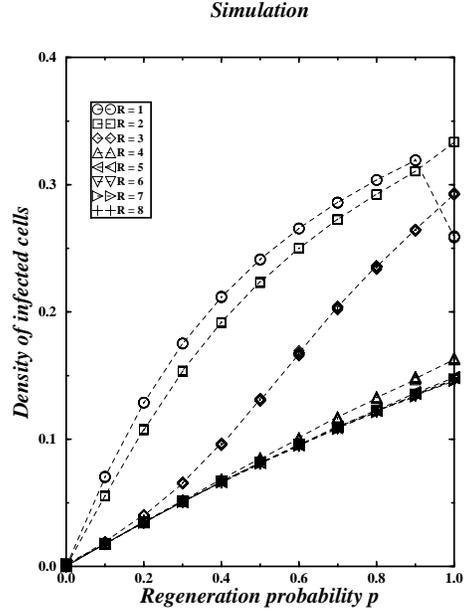}}
\caption{Mean density of infected cells $D_i$ obtained by simulation for a
$L=100$ square lattice as function of the regeneration probability $p$.
$R=1-8$, $f/p=0.1$, $q=0.6$. Dashed lines are guide for the eyes.}
\label{fig1}
\end{center}
\end{figure}


\begin{figure}
\begin{center}
\leavevmode
\vbox{%
\epsfxsize=6cm
\epsffile{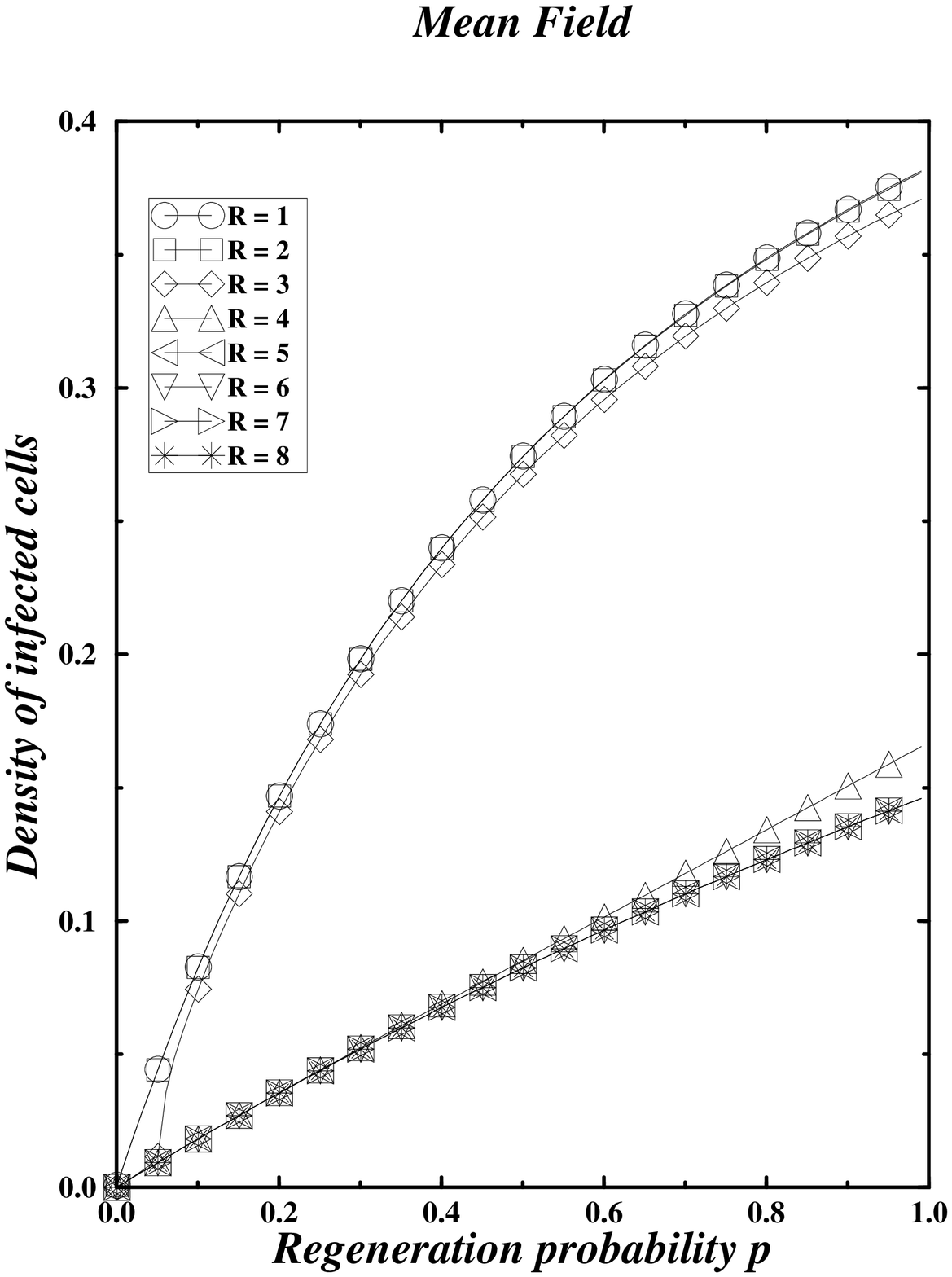}}
\caption{Mean density of infected cells $D_{i}$ obtained by mean field
approximation, as function of the regeneration probability $p$. Same parameters
of figure 1.}
\label{fig2}
\end{center}
\end{figure}


\begin{figure}
\begin{center}
\leavevmode
\vbox{%
\epsfxsize=6cm\epsffile{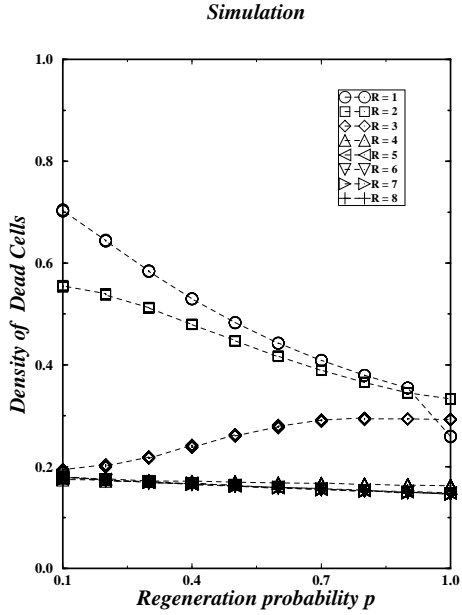}}
\caption{Mean density of dead cells $D_{d}$ obtained by simulation on a $L=100$
square lattice as function of the regeneration probability $p$. Same parameters
of figure 1.}
\label{fig3}
\end{center}
\end{figure}


\begin{figure}
\begin{center}
\leavevmode
\vbox{%
\epsfxsize=6cm\epsffile{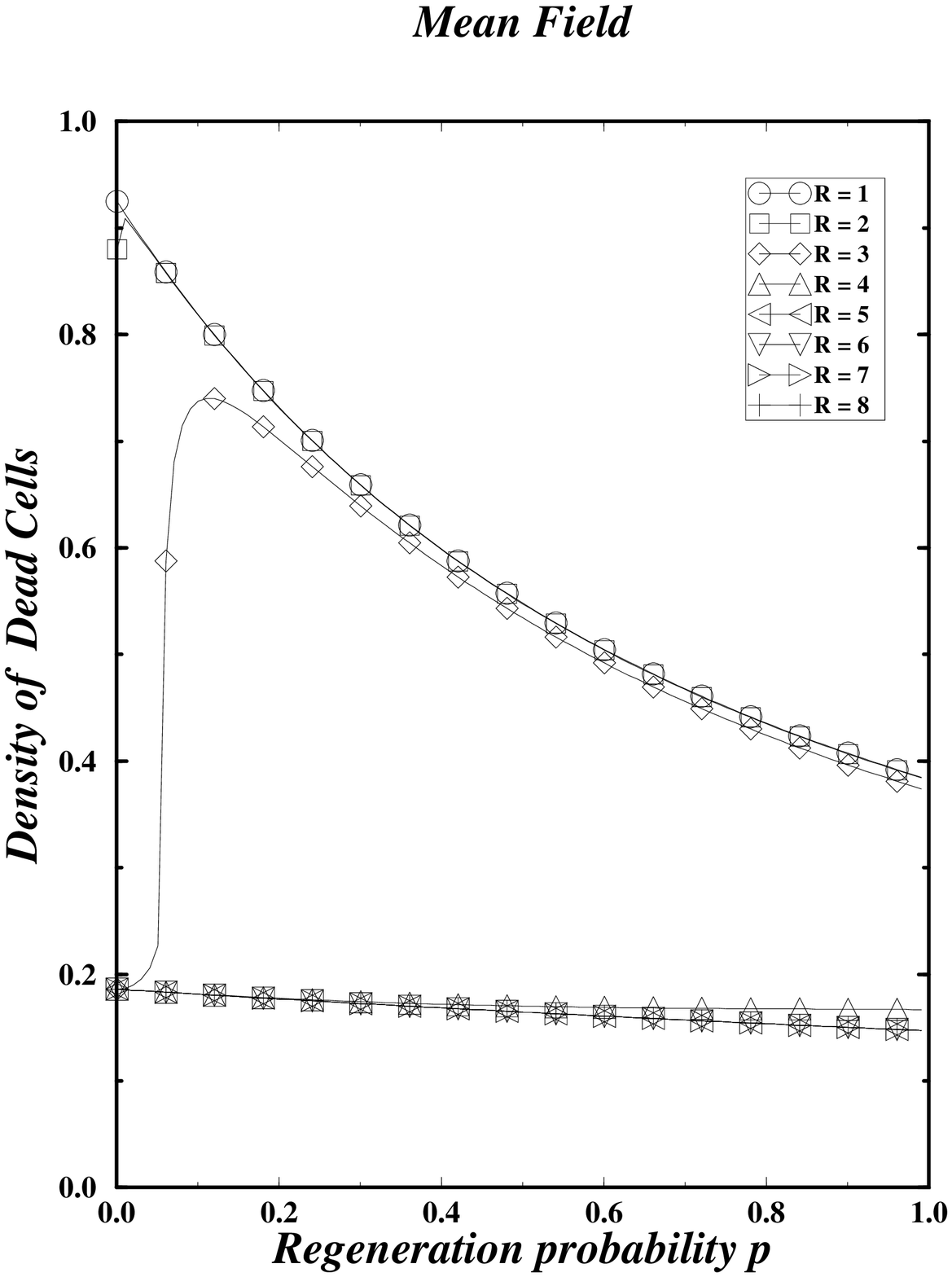}}
\caption{Mean density of dead cells obtained by mean field approximation, as
function of the regeneration probability. Same parameters of figure 1.}
\label{fig4}
\end{center}
\end{figure}


\begin{figure}
\begin{center}
\leavevmode
\vbox{%
\epsfxsize=6cm\epsffile{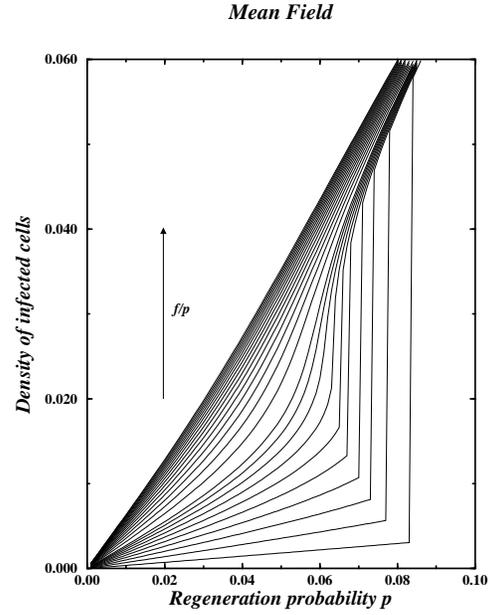}}
\caption{Mean density of infected cells $D_i$ obtained by the mean field
approach as function of the regeneration probability $p$ for $R=3$ for several
values from the ratio $f/p=0.0001$ to $1.0$,  with $q=0.6$.}
\label{fig5}
\end{center}
\end{figure}


\begin{figure}
\begin{center}
\leavevmode
\vbox{%
\epsfxsize=6cm\epsffile{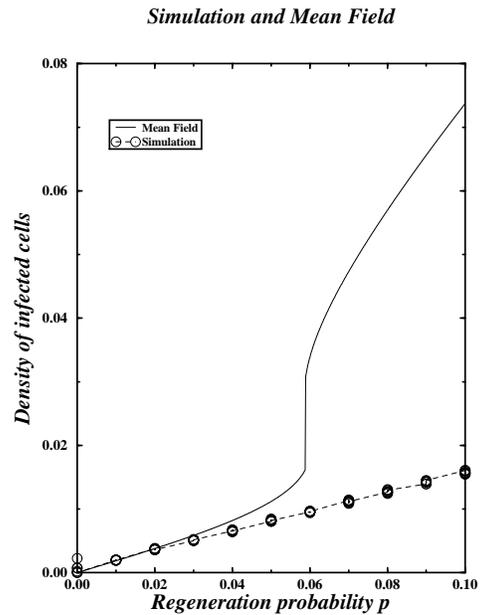}}
\caption{Mean density of infected cells $D_i$ as function of the regeneration
probability $p$ for resistance parameter $R=3$. Full line is obtained from mean
field approach and open circles indicates figures from simulations. Same
parameters of figure 1.}
\label{fig6}
\end{center}
\end{figure}


\begin{figure}
\begin{center}
\leavevmode
\vbox{%
\epsfxsize=6cm
\epsffile{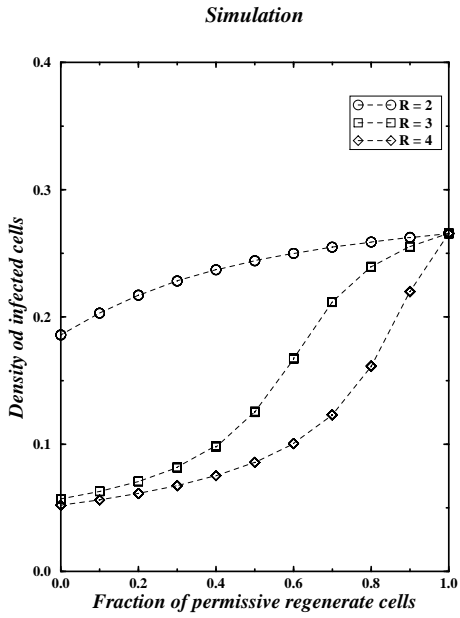}}
\caption{Mean density of infected cells $D_i$ obtained by simulation on a
$L=200$ square lattice as function of the fraction of regeneration of
permissive cells $q$ for $f/p=0.1$ and $p=0.6$.}
\label{fig7}
\end{center}
\end{figure}


\begin{figure}
\begin{center}
\leavevmode
\vbox{%
\epsfxsize=6cm
\epsffile{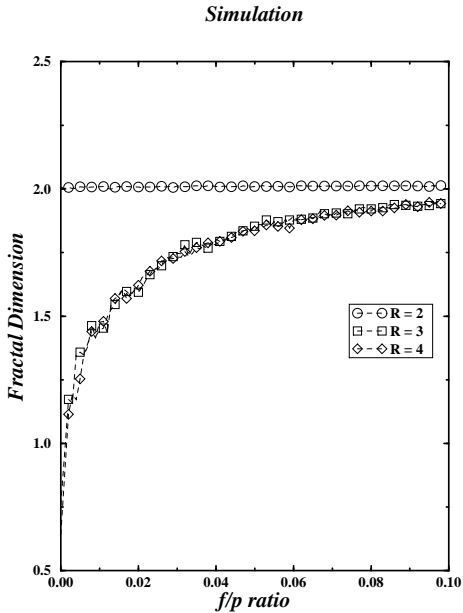}}
\caption{Fractal dimension of ulcers contours $D_{F}$
 obtained by simulation as function of the $f/p$ ratio
 for $p=0.01$ and $q=0.6$ on a $L=100$ square lattice,
 for $R = 2,3$ and $4$.}
\label{fig8}
\end{center}
\end{figure}

\end{document}